\documentclass[A4]{elsart}

\usepackage{epsfig}
\usepackage{psfrag}
\usepackage{amssymb}
\begin{document}

\begin{frontmatter}

\title{Sedimentation of Oblate Ellipsoids at low and Moderate Reynolds numbers}
\author{F. Fonseca and H. J. Herrmann}
\address{ ICA-1, University of Stuttgart, Pfaffenwaldring 27, 70569
             Stuttgart,  Germany}

\begin{abstract}
In many applications to biophysics and environmental engineering, sedimentation
of non-spherical particles for example: ellipsoids, is an important
problem. In our work, we simulate the dynamics of oblate ellipsoids under gravity. We study the settling velocity and the average orientation of the ellipsoids
as a function of volume fraction. We see that the settling velocity shows a local maximum at the intermmediate densities unlike the spheres. The 
average orientation of the ellipsoids also shows a similar local maximum and we
observe that this local maximum disappears as the Reynolds number is 
increased. Also, at small volume fractions, we observe that the oblate
ellipsoids exhibit an orientational clustering effect in alignment with
gravity accompanied by strong density fluctuations. The vertical
and horizontal fluctuations of the oblate ellipsoids are small compared
to that of the spheres.
\end{abstract}
\end{frontmatter}

\section{Introduction}

The sedimentation of particles in a fluid under the action of gravity is an 
interesting problem in fluid dynamics and statistical physics 
which finds application in biology,
environmental and industrial engineering. For example, chemical reactors,  
pollution, ink-jet printing, blood, fluidized beds, etc.  
This problem reveals complicated multi-body interactions 
due to the long-range hydrodynamics, 
that decay for a sphere as $1/r$, where $r$ is 
the distance between the particles.
The particle velocities strongly fluctuate and 
there has been some controversy about the nature 
of these velocity fluctuations, \cite{Ramas}.

In the dilute case and low Reynolds numbers, 
the behavior of the settling velocity $V(\phi)$ as 
a function of the volume fraction $\Phi$, in the limit where
Brownian motion is negligible, has been investigated by
Batchelor\cite{Batche}. 
 $V(\Phi)$ decreases monotonically as $\Phi$ 
increases following the phenomenological Richardson-Zaki law \cite{RiZaki}.

Much of the investigations have been made on the 
spheres and in a reduced manner on 
slender bodies, e.g. fibers by \cite{Ramas,Batche1}.
Fibers have many technological applications\cite{paper}.
An orientational transition is well known for smaller
volumer fractions, which is characterized by
 a maximum in the sedimentation velocity \cite{Esa1}.
Ellipsoidal particles find applications in the models
for blood flow \cite{Olla}, car paint and sun protectors.

In our previous work \cite{FonsH}, we have constructed a phase-diagram
where three basic regimes exist namely 
steady-falling, oscillatory and chaotic regime for a single oblate.
In this paper, we present results of the simulation 
of many oblates sedimenting in the fluid 
under gravity, in the steady falling regime of a single oblate.
The rest of the paper is organized as follows. 
In section 2, we discuss the model, in section 3.1, 
we present the behavior of sedimentation velocity, 
in 3.2, we discuss the orientation of  
ellipsoids, in 3.3, we discuss orientational order parameter. In section
3.4, we present results at moderate Reynolds number with not 
very large aspect ratios followed by
summary and conclusion in section 4.


\section{Model}

\begin{figure}
\begin{center}
\epsfig{file=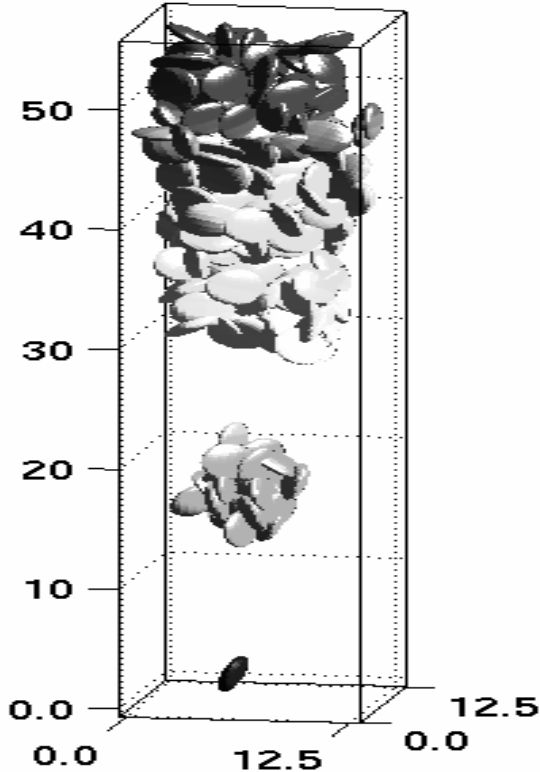,angle=0,width=8cm,height=11cm}
\caption{Snapshot of the oblate ellipsoids falling in a fluid. 
The picture shows the 
``cluster'' formation. The ellipsoid aspect-ratio is $\Delta r=0.4/1.5$, 
the distance is given in units of the 
larger radius $R_{M}$, the Reynolds number is 
$Re=4*10^{-2}$ and the particle volume fraction is $\Phi_{V}=0.05$.}
\label{fig1} 
\end{center}
\end{figure}
 
The model was developed by H\"{o}efler and Schwarzer \cite{Kai,KS}, 
extended by Kuusela et al \cite{Esa} 
and has been applied to several studies 
\cite{Esaprl,FonsH}.
The fluid motion is solved by discretizing on a mesh \cite{KS}, the incompressible Navier-Stokes 
equations:
\begin{equation}
\label{NavierStokesEquations}
\begin{array}{c}
\frac{\partial\vec{v}}{\partial t}+(\vec{v}\cdot \vec{\nabla })\vec{v}=-\vec{\nabla }p+\frac{1}{{Re}}\nabla ^{2}\vec{v}+\vec{f} \\
\vec{\nabla }\cdot \vec{v}=0
\end{array}
\end{equation}
where $\vec{v}$ is the fluid velocity, $p$ the pressure and $\vec{f}$
 represents an external force, which in our problem is gravity. 
The boundary conditions between the fluid 
and the oblate ellipsoid particles are satisfied 
considering that the fluid motion on the particle surface is subject to the non-slip boundary
condition:
\begin{equation}
\vec{v}(\vec{x})= \vec{v_{t}} + \vec{r}(\vec{x})_{CM}\times\vec{\omega}
\end{equation}
\noindent where $\vec{v}_{t}$ is the translational velocity of the ellipsoid, $\vec{r}(\vec{x}_{CM})$ the
vector from its center to the point $\vec{x}$ at the ellipsoid surface, and $\vec{\omega}$ the ellipsoid 
angular velocity.
The interaction between the ellipsoidal surface and the fluid adjacent
to the ellipsoid is achieved by setting a restoring force that gives rise
to a "distribution force" in the body term of the Navier Stokes equation.
This distribution force mimics the presence of the ellipsoids in the sense
that the fluid inside the ellipsoid moves like a rigid body. A restoring 
force is applied whenever the template particle and the fluid rigid body are not in the same position 
\cite{KS}.  

The repulsive force between the ellipsoids is chosen proportional to their overlap.
When the oblate ellipsoids are non-overlapping the force is zero and 
at short distances 
the hydrodynamic forces describing the existence of 
the fluid avoid the contact between the 
particles \cite{Esa}. For this force, we choose a 
contact function explained in 
depth in \cite{Esa} and \cite{Perram} .
The geometry of the oblate ellipsoid is 
characterized by $\Delta r$, its aspect-ratio is  
defined as the ratio of the minimum radius $R_{m}$ 
to the maximum radius, $R_{M}$:
\begin{equation}
\Delta r= \frac{R_{m}}{R_{M}}  
\end{equation}

We define the equivalence between the sphere and an oblate ellipsoid 
as the sphere that has the same 
volume, with the equivalent radius:
\begin{equation}
R_{equi}= \sqrt[3]{R_{m}R^{2}_{M}}  
\end{equation} 
The Reynolds number is defined as:
 \begin{equation}
Re:=\frac{2R_{M}v_{s}\rho_{f}}{\nu}
\end{equation}
\noindent where $v_{s}$ is the mean vertical oblate 
ellipsoid velocity, $2*R_{M}$ being the characteristic 
length,
 $\rho_{f}$ is the density 
and $\nu$ is the dynamical viscosity of the fluid. 
The number of ellipsoids in our simulations is of the order of one thousand.
We choose the density of the fluid, the Stokes velocity 
and the larger radius of the 
ellipsoid equal to unity in our system. In all cases the container
 has a height, $L=85$ 
and a base of $12.5\times12.5$, and the lattice constant, $h=0.7$. The ratio 
between the density of the oblate ellipsoids and the fluid is $4$.


\section{Results}
\subsection{Sedimentation velocity for oblate ellipsoids}
\begin{figure}
\begin{center}
\epsfig{file=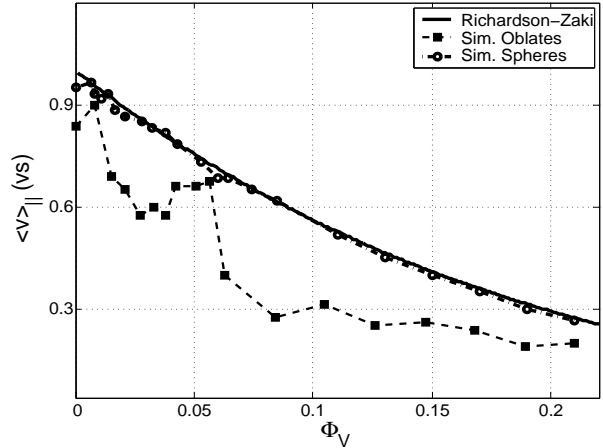,angle=0,width=8cm,height=6cm}
\caption{Mean sedimentation velocity $v(t)_{\parallel}$ for 
the oblate ellipsoid (dash-squared 
line) and sphere (dash-dot line), as function of the volume fraction $\Phi_{V}$. The oblate 
ellipsoid aspect-ratio is $\Delta r=0.4/1.5$, the radius of the 
equivalent sphere $R_{equi}=0.97$ and the 
Reynolds number $Re=4*10^{-2}$.}
\label{fig2} 
\end{center}
\end{figure}

In fig. \ref{fig2} we show the mean vertical sedimentation velocity $v(t)_{\parallel}$ 
as a function of the volume fraction $\Phi_{V}$, within a range 
of 0.01 to 0.21, for 
oblate ellipsoids and spheres and then compared to the 
phenomenological Richardson-Zaki law 
$\frac{V(\Phi)}{V_{0}}=(1-\Phi_{V})^{n}$ \cite{RiZaki} with $n=5.5$. 
The limit of
$\Phi_{V}{\rightarrow 0}$ corresponds to the single 
falling ellipsoid which we studied in our previous work
\cite{FonsH}. It is interesting to point out that 
the sedimentation velocity of the ellipsoid, is small compared to that
of the sphere, which follows the 
phenomenological Richardson-Zaki law. 
This is not the case for fibers (elongated ellipsoids), 
where it is found that the sedimentation velocity has a maximum 
for smaller volume fraction which 
can exceed the terminal velocity of a single fiber \cite{Esaprl}.

For oblate ellipsoids the mean vertical sedimentation velocity 
passes through a local 
maximum at $\Phi_{V}\approx0.05$. This maximum is quite
 interesting since it is not observed 
for spheres. Similar non-monotonic sedimentation of 
non-spherical bodies (e. g. fibers) has 
been reported experimentally by Herzhaft et. al.\cite{Herzhaft} and for 
prolate ellipsoids in  
simulations by Kuusela et al. 
\cite{Esa} due to an orientation parallel to gravity.

\begin{figure}
\begin{center}
\epsfig{file=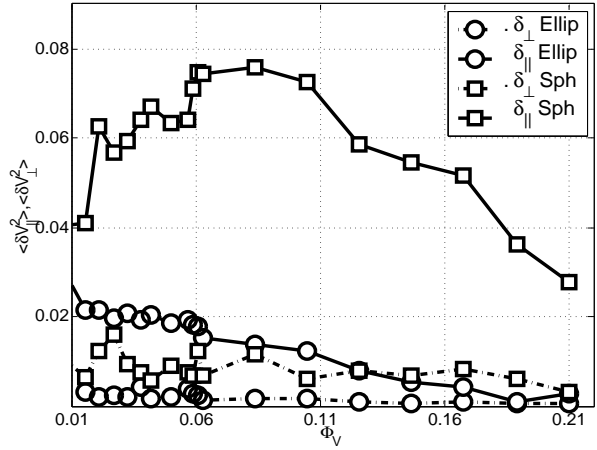,angle=0,width=8cm,height=6cm}
\caption{Velocity fluctuations for ellipsoids (circle-line) 
and spheres (squared-line) 
for the vertical (solid line), and horizontal (dash-dot line) 
components corresponding to 
fig. $1$. The oblate ellipsoid has 
an aspect-ratio of $\Delta r=0.4/1.5$, the equivalent 
radius of a sphere is $R_{equi}=0.97$, and the Reynolds number $Re=4*10^{-2}$.}
\label{fig3} 
\end{center}
\end{figure}

In fig. \ref{fig3} we present the parallel ($\parallel$) and perpendicular ($\bot$) 
components of the velocity fluctuations with respect to gravity 
as a function of the volume fraction
$\Phi_{V}$ which are defined as:
\begin{equation}
\delta V^{2}_{\parallel}=<V^{2}_{\parallel}>-<V_{\parallel}>^{2}
\end{equation}
\begin{equation}
\delta V^{2}_{\bot}=<V^{2}_{\bot}>
\end{equation}
The angular brackets indicate averaging over the ellipsoids that have not reached 
the final bottom position at the container. The averages were made over at least 50 
realizations starting with different random postions and orientations. In fig. 
\ref{fig3} the vertical (parallel to gravity) fluctuations for spheres and for ellipsoids 
are much larger than the 
respective horizontal components. The fluctuations for ellipsoids 
decrease with the 
volume fraction. For an equivalent system of spheres, 
the fluctuations show a maximum at 
intermediate volume fractions ($\Phi_{V}\approx0.07 $) \cite{Kalthoff,Nicolai}. 
In all cases the fluctuations for the spheres are considerably larger than the 
fluctuations for oblate ellipsoids.
\begin{figure}
\begin{center}
\epsfig{file=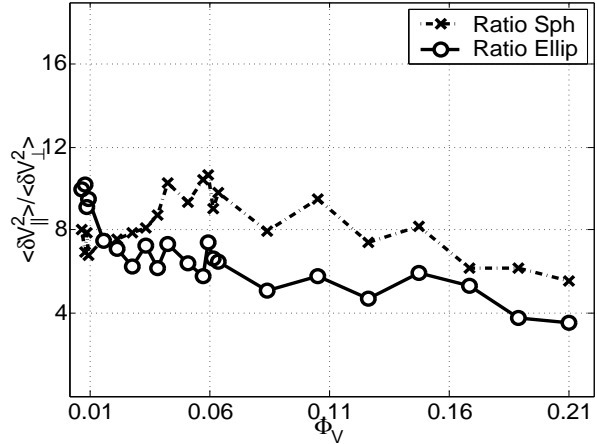,angle=0,width=8cm,height=6cm}
\caption{Ratio of the vertical to the horizontal velocity 
fluctuations for spheres and 
oblate ellipsoids as function of the 
volume fraction, $\Phi_{V}$. The oblate 
ellipsoid aspect-ratio is $\Delta r=0.4/1.5$, 
and the Reynolds number $Re=4*10^{-2}$.}
\label{fig4} 
\end{center}
\end{figure}

In fig. \ref{fig4} we present the ratio, $\delta V_{||}^{2}/\delta V_{\perp}^{2}$ for 
spheres and oblate ellipsoids. For spheres the ratio shows a maximum around 
$\Phi_{V}\approx0.07$ \cite{Kalthoff,Nicolai}. For ellipsoids the ratio 
shows a slightly larger value than that of the spheres 
for very small volume fractions and 
has an overall monotonic decrease with the volume fraction.  
\begin{figure}
\begin{center}
\epsfig{file=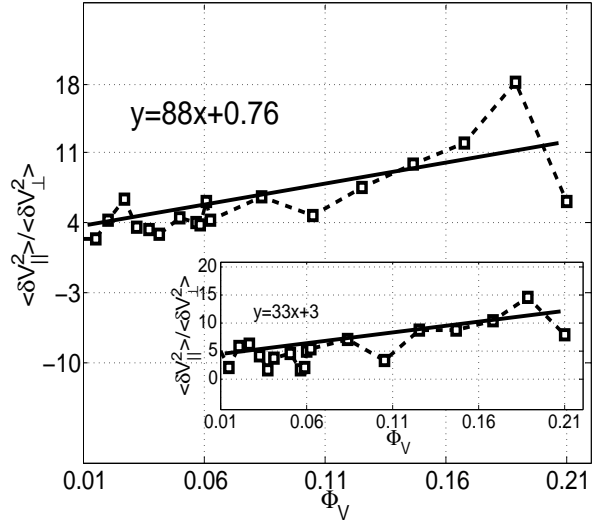,angle=0,width=8cm,height=7cm}
\caption{Ratio of the vertical velocity fluctuations for spheres to that of
the oblate ellipsoids as function of the volume fraction, $\Phi_{V}$. 
The inset shows the 
corresponding ratio for the horizontal fluctuations. 
The oblate ellipsoid aspect-ratio 
is $\Delta r=0.4/1.5$, and the Reynolds number $Re=4*10^{-2}$.}
\label{fig5} 
\end{center}
\end{figure}

We display the ratio of the vertical velocity fluctuations for spheres to 
that of the ellipsoids in fig. \ref{fig5}. The quotient 
exhibits a linear behavior 
with the volume fraction following approximately the relation 
$(\delta V_{vert,sph}^{2}/\delta V_{vert,ellip}^{2})=88*\Phi_{V}+0.76$.
The inset shows the horizontal case, also a linear behavior
$(\delta V_{hor,sph}^{2}/\delta V_{hor,ellip}^{2})=33*\Phi_{V}+3$.
\begin{figure}
\begin{center}
\epsfig{file=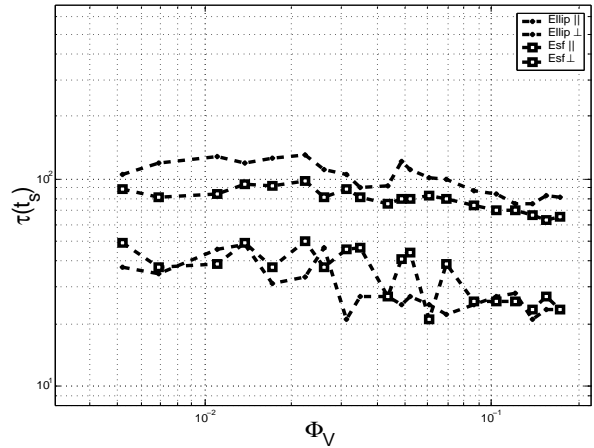,angle=0,width=8cm,height=6cm}
\caption{ Autocorrelation times $\tau$ (in units of $t_{s}$) as function 
of the volume fraction $\Phi_{V}$ for oblate ellipsoids 
and the equivalent spheres split 
into components parallel ($||$) and 
perpendicular ($\perp$) to gravity. The oblate 
ellipsoid aspect-ratio is $\Delta r=0.4/1.5$, 
the equivalent sphere system has a radius 
$R_{equi}=0.97$ and the $Re=4*10^{-2}$. 
The results are plotted in a log-log scale.}
\label{fig6} 
\end{center}
\end{figure}

In figure \ref{fig6} we present the parallel, $\tau_{\parallel}$
and perpendicular, $\tau_{\perp}$ components of the autocorrelation time for
 both ellipsoids and spheres. 
 We use the definition of the correlation time as:
\begin{equation}
\tau_{c}=\frac{1}{C(0)}{\int_0^\infty C(t)dt}
\end{equation}
Where $C(t)$ is the particle velocity autocorrelation 
function which is defined as 
$C(t)=<\delta V(t)\delta V(0)>$. Here $\delta V(t) = V(t)- <V>$ 
is the local velocity 
fluctuation, where $<V>$ was taken as the mean 
(horizontal or vertical) velocity.

The vertical component, shows an overall large value for the 
oblate ellipsoids compared to that of the spheres. 
The horizontal components of the autocorrelation time
 between oblate ellipsoids 
and spheres are indistiguishable.

In figure \ref{fig6} the four curves decay as power laws given by
 $\tau\approx\Phi^{-\alpha}$,
We see that the values of $\alpha$ for the parallel 
and perpendicular components for ellipsoids are 
$\alpha_{\parallel}\approx 0.11$ and $\alpha_{\perp}\approx 0.16$ respectively
and that for the spheres are $\alpha_{\parallel} \approx -0.10$ and 
$\alpha_{\perp}\approx 0.22$ respectively

\begin{figure}
\begin{center}
\epsfig{file=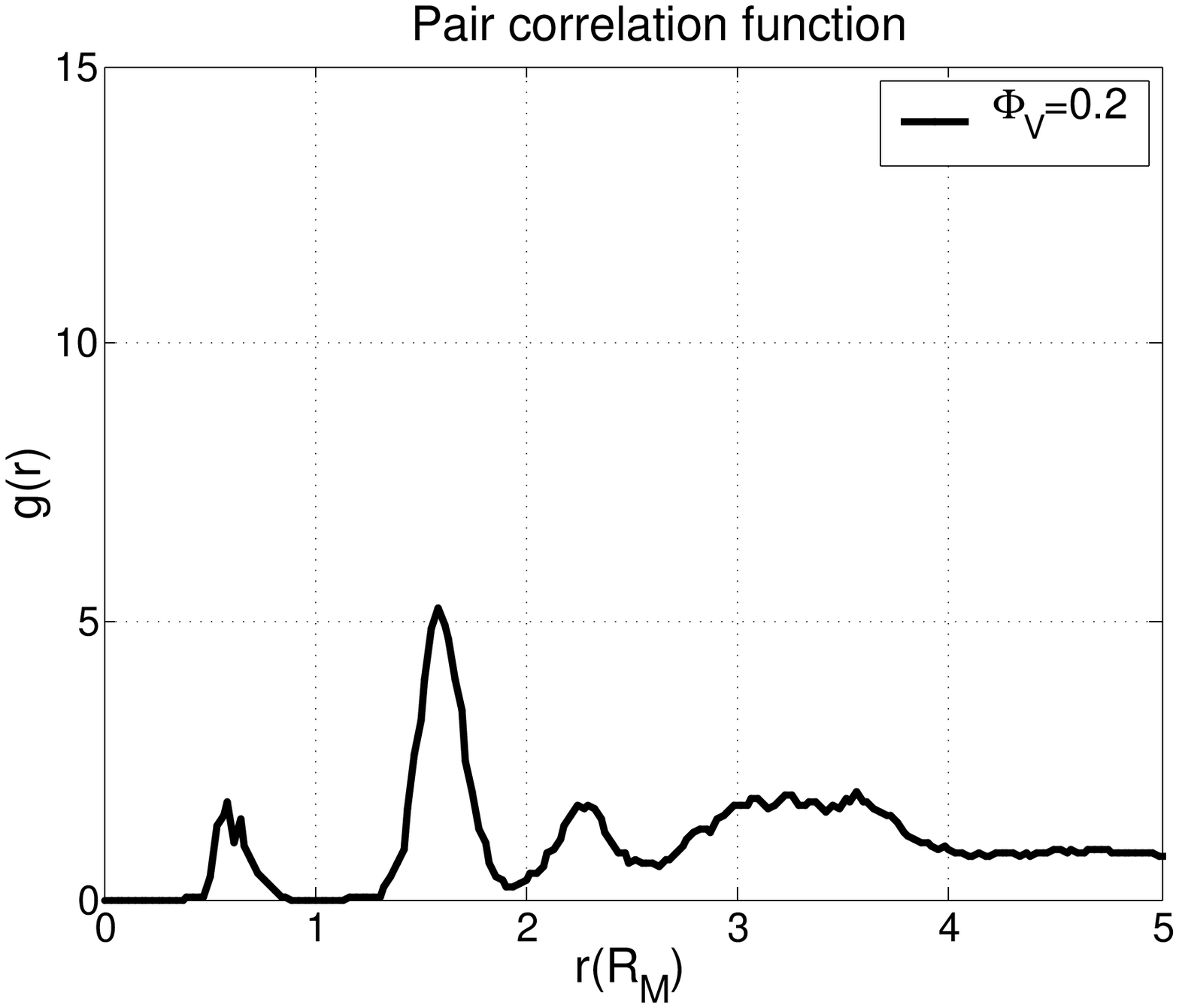,angle=0,width=7cm,height=5cm}
\epsfig{file=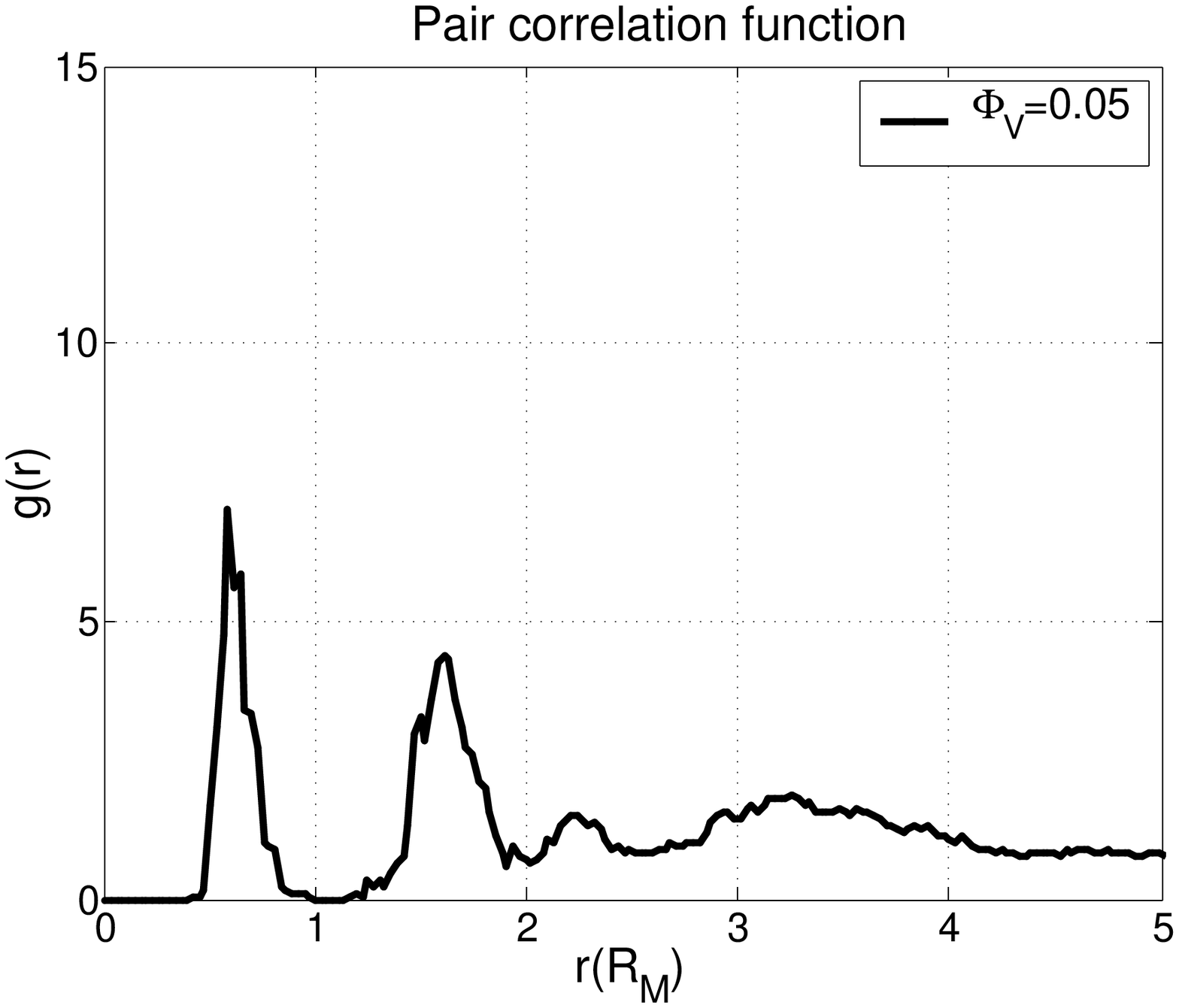,angle=0,width=7cm,height=5cm}
\epsfig{file=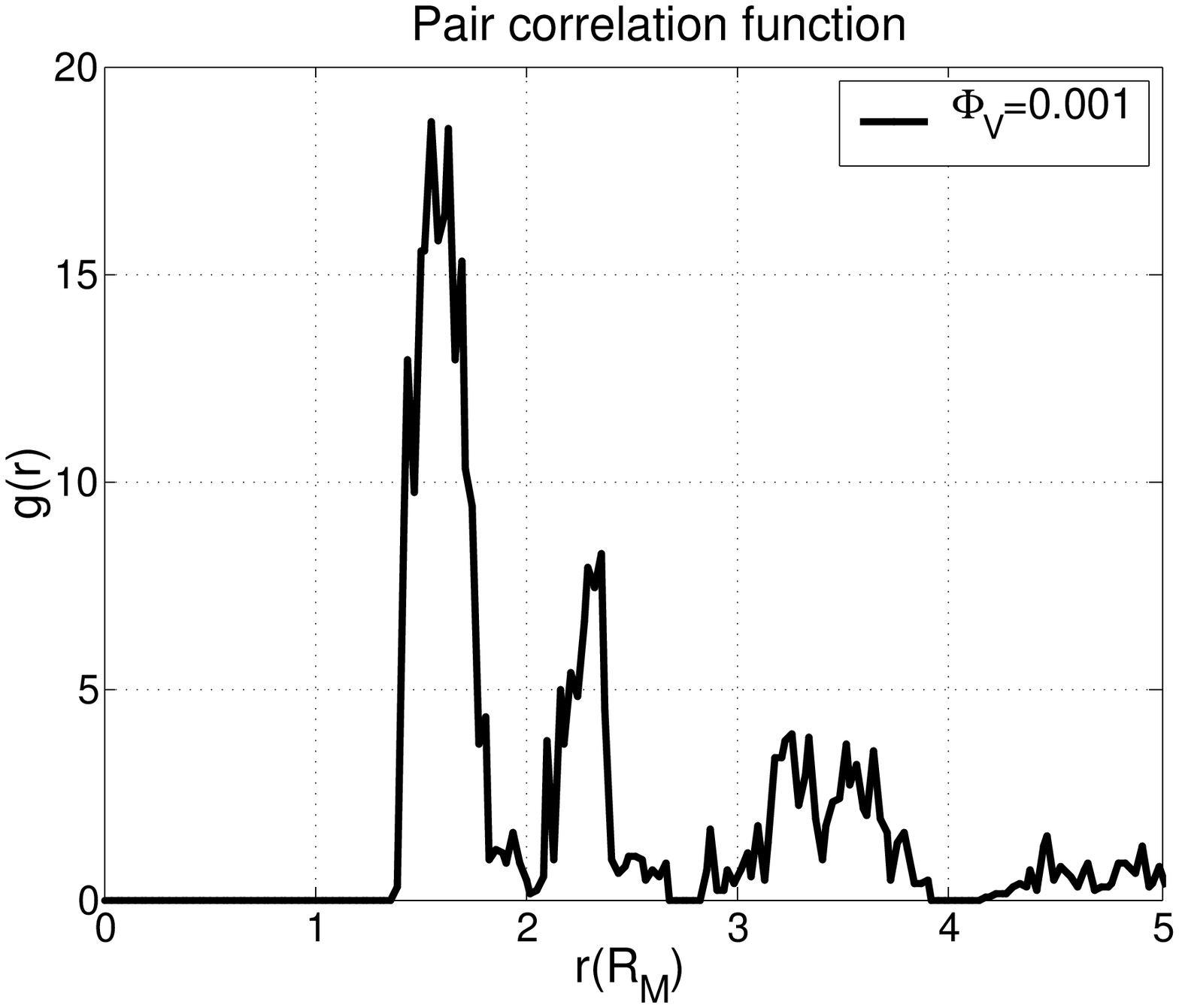,angle=0,width=7cm,height=5cm}
\caption{Pair distribution functions for 
oblate ellipsoids for different volume fractions, 
$\Phi_{V}$. The Reynolds number $Re=4*10^{-2}$.}
\label{fig7} 
\end{center}
\end{figure}

We calculate the pair correlation function for different volume 
fractions $\Phi_{V}$ and the results are shown in fig. \ref{fig7}. The pair 
correlation function, for smaller volume fractions $\Phi_{V}=0.001$, 
clearly shows large inhomogeneities in the sense that there 
is a ``packing formation'' 
as seen in fig. \ref{fig1}, of oblate ellipsoids. These inhomogeneities 
disappear for large volume fraction $\Phi_{V}\geq 0,2 $. 
Furthermore, in the intermediate case for $\Phi_{V}=0.05$ 
we can see in the pair 
correlation function that the first peak is close 
to the origin, located at $r=0.6$, which 
is also present at $\Phi_{V}=0.2$ but smaller. This additional larger peak at 
$\Phi_{V}=0.05$ could be related to the local maximum 
in the sedimentation velocity fig. 
\ref{fig2}. By looking at the snapshots as the 
one shown in fig. \ref{fig1} one sees that 
entire bundles of aligned particles seem to detach 
and move down faster which might well
be the origin of this peak. This kind of ``bundle behavior'' has also been 
observed in the 
sedimentation of fibers ref \cite{Herzhaft} where these bundles settle faster than the 
individual fibers.

\subsection{Orientational behavior}

\begin{figure}
\begin{center}
\epsfig{file=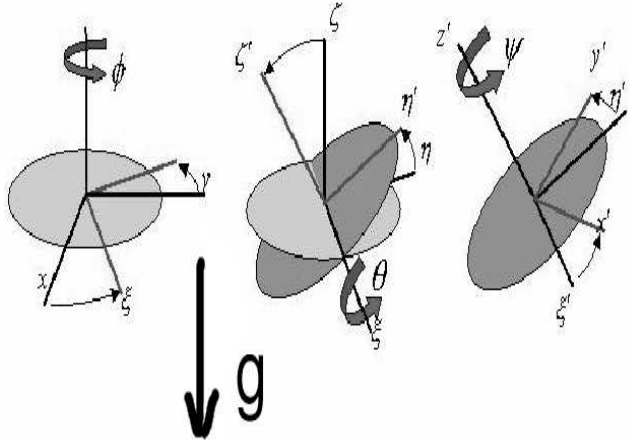,angle=0,width=8.5cm,height=6cm}
\caption{Euler angles $\phi$, $\theta$ and $\psi$ 
used for the description of the 
oblate ellipsoid orientational behavior $[17]$.}
\label{fig8} 
\end{center}
\end{figure}

\begin{figure}
\begin{center}
\epsfig{file=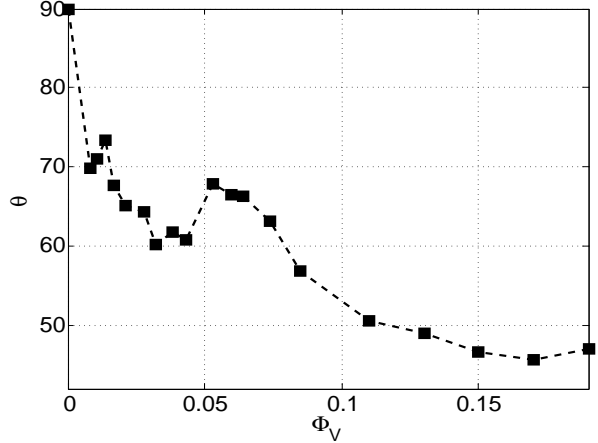,angle=0,width=8cm,height=6cm}
\caption{Mean vertical orientation $\theta$  for oblate ellipsoids 
as a function of the 
volume fraction $\Phi_{V}$. The oblate ellipsoid 
aspect-ratio is $\Delta r=0.4/1.5$, and the 
Reynolds number $Re=4*10^{-2}$.}
\label{fig9} 
\end{center}
\end{figure}

For the measurement of the orientation we use the 
Euler angles described in fig. \ref{fig8}. 
The mean vertical orientation (MVO), $\theta$ , 
as a function of the volume fraction, is 
shown in fig. \ref{fig9}. For smaller volume fraction 
the MVO shows more alignment 
with gravity and in the limit $\Phi_{V}\rightarrow 0$
a closer alignment with gravity is observed
which corresponds to the orientational behavior 
for one oblate ellipsoid observed 
in \cite{FonsH}. We also see for 
the MVO an intermediate maximum, at $\Phi_{V}\approx0.05$, 
which could explain the local 
vertical velocity maximum at the same volume fraction 
shown in fig. \ref{fig2}.  
This intermediate maximum is not present for spheres. 
For larger values of the volume fraction, 
$\Phi_{V}>0.08$ the plot shows a monotonic decrease.

\begin{figure}
\begin{center}
\epsfig{file=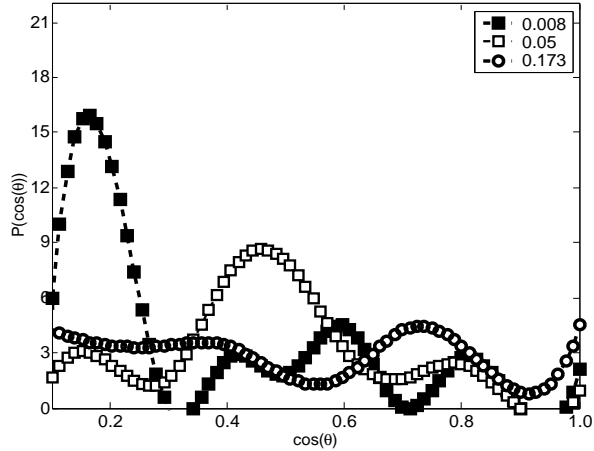,angle=0,width=8cm,height=6cm}
\caption{The distribution function $P(cos(\theta))$ 
for the mean vertical orientation 
$\theta$  for different volume fractions. 
The ellipsoid aspect-ratio is $\Delta r=0.4/1.5$, and the 
Reynolds number $Re=4*10^{-2}$.}
 \label{fig10} 
\end{center}
\end{figure}

Figure \ref{fig10} shows the orientational distribution 
function $P(cos(\theta))$ 
for the vertical angle, $\theta$, for different 
volume fractions, $\Phi_{V}$. For smaller 
volume fractions, $\Phi_{V}=0.008$ the orientational 
distribution shows a maximum around 
$cos(\theta)\approx 0.1$ in agreement with fig. \ref{fig9}. 
The limiting case ($\Phi_{V} \rightarrow 0$), i.e., 
one sedimenting oblate ellipsoid, studied by 
us in \cite{FonsH},  
presents a vertical alignment 
with gravity ($\theta \approx 90^{o}$), and in 
fig \ref{fig9} we can see a value of $\theta \approx 85^{o}$. 

As the volume fraction increases, the distributions 
flatten, and at $\Phi_{V}=0.05$ the 
distribution shows a moderate maximum around $cos(\theta)\approx 0.45$, 
corresponding to the similar intermediate 
maximum in figures \ref{fig9} and \ref{fig2}. 
We conclude that the vertical velocity is influenced 
significatively by the orientational 
behavior along gravity, as it is well known for 
other spheroid systems \cite{Esa1}.

\begin{figure}
\begin{center}
\epsfig{file=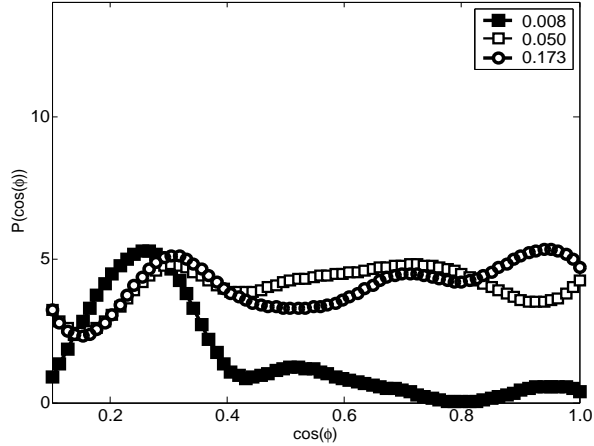,angle=0,width=8cm,height=6cm}
\caption{The distribution function $P(cos(\phi))$ for the mean vertical 
orientation $\phi$ for oblate ellipsoids. 
The aspect-ratio is $\Delta r=0.4/1.5$, and the 
Reynolds number $Re=4*10^{-2}$.  }
\label{fig12} 
\end{center}
\end{figure}

Figure \ref{fig12} shows the orientational distribution function $P(cos(\phi))$ 
for the angle $\phi$, for different volume fractions, $\Phi_{V}$. 
The orientation around
the vertical slightly increases for smaller volume fractions, 
and decreases with larger volume 
fractions. Similar behavior is also found for 
the third Euler angle $\psi$. We conclude 
that the Euler angles $\phi$ and $\psi$ are not much 
influenced by the volume fraction.

\subsection{Orientational changes}

\begin{figure}
\begin{center}
\epsfig{file=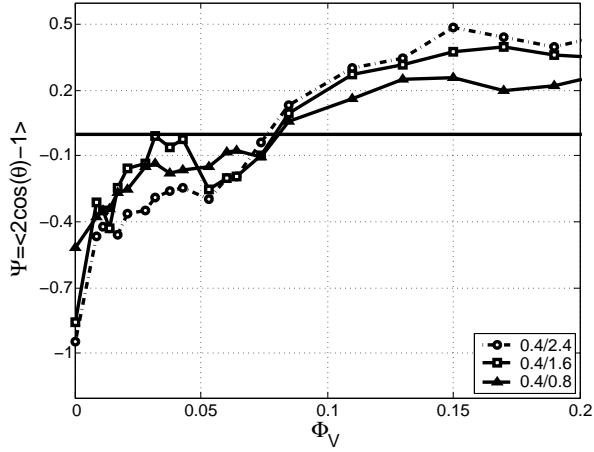,angle=0,width=8cm,height=6cm}
\caption{Order parameter $\Psi$ as a function of 
the volume fraction, $\Phi_{V}$ for three
different aspect-ratios $\Delta r=0.4/2.4$; $0.4/1.6$; $0.4/0.8$.}
\label{fig15} 
\end{center}
\end{figure}

To quantify the orientation of the oblate ellipsoids we introduce the quantity 
$\Psi=<2cos(\theta)-1>$ that was also used in \cite{Esa1}, \cite{Herzhaft} as 
orientational order parameter, it would 
give $-1$, $0$ or $+1$ if all the oblate ellipsoids
were perpendicular to gravity, randomly oriented or aligned with gravity
respectively. Figure 
\ref{fig15} shows the behavior of $\Psi$ 
against $\Phi_{V}$, for smaller volume fractions, 
$\Phi_{V}\approx0.001-0.08$ the order parameter 
takes negative values evidencing the alignment
along gravity and in agreement with the limit,  
$\Phi_{V}\rightarrow 0$ (one oblate ellipsoid)
\cite{FonsH}. 
Approximately at $\Phi_{V}\approx0.08$ the 
order parameter is zero.
For larger $\Phi_{V}\geq0.08$ 
a positive order parameter implies the orientation 
is perpendicular to gravity.  

In the range of $\Phi_{V}\approx0.001-0.08$, 
$\Psi$ has a local minimum close to 
$\Phi_{V}\approx0.05$ where we found a 
local maximum in fig. \ref{fig9} and fig. \ref{fig2}.  
The simulations were repeated with two other 
different aspects ratios $A_{r}=0.4/0.8; 0.4/2.4$ 
and we observed similar behavior. In the case 
of one oblate ellipsoid ($\Phi_{V}\rightarrow 0$) 
the order parameter $\Psi$ has a value very close to
$-1$ as the ellipsoid aspect-ratio is increased \cite{FonsH}.

\subsection{Moderate Reynolds number}

\begin{figure}
\begin{center}
\epsfig{file=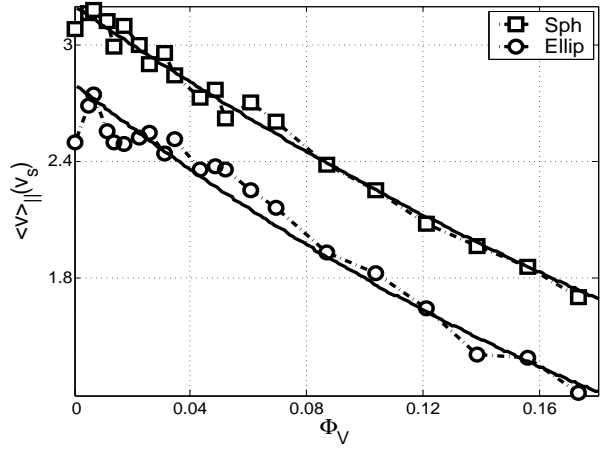,angle=0,width=8cm,height=6cm}
\caption{Mean sedimentation velocity $v(t)_{\parallel}$ 
for the oblate ellipsoid (dash-squared 
line) and a sphere (dash-circle line), 
as function of the volume fraction, $\Phi_{V}$. The oblate 
ellipsoid aspect-ratio is $\Delta r=0.4/1.5$, 
the equivalent sphere has $R_{equi}=0.97$ and the 
Reynolds number $Re\approx 7$.}
\label{fig16} 
\end{center}
\end{figure}

Figure \ref{fig16} presents the mean vertical 
sedimentation velocity for oblate ellipsoids 
($\square$ squared line) and the 
equivalent spheres ($\circ$ circle lined) as a function of the 
volume fraction at moderate Reynolds number ($Re\approx$ 7). 
In our previous work this simulation method has been used 
with success up to $Re\approx 10$ \cite{KS} 
and \cite{Esaprl}. The intermediate maximum for
the ellipsoids is not observed in fig. \ref{fig16}  
as seen in fig. \ref{fig2} at low Reynolds number.

A comparison with the phenomenological 
Richardson-Zaki law (continous line in fig \ref{fig16}) shows 
an exponent around $n_{Sph}=3.2$ for spheres 
and $n_{Ellip}=4.0$ for ellipsoids. In both both cases, 
the data follow the Richardson-Zaki law 
rather closely. These exponents ($n_{Sph}=3.2$ and 
$n_{Ellip}=4.0$) are between the low particle 
Reynolds number limit ($n\approx 4.5$) and a turbulent 
particle system ($n\approx 2.5$), \cite{RiZaki}.

\begin{figure}
\begin{center}
\epsfig{file=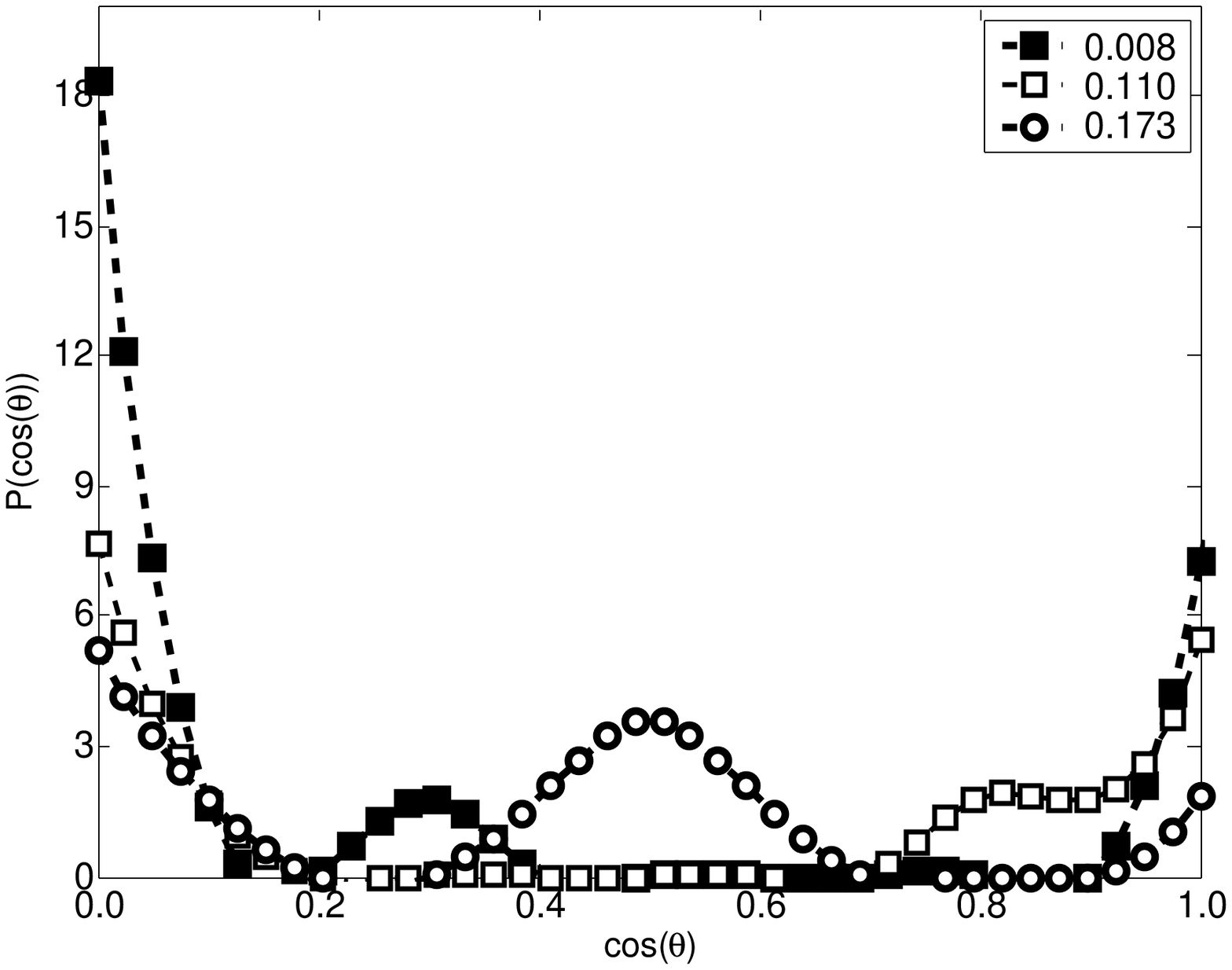,angle=0,width=8cm,height=6cm}
\epsfig{file=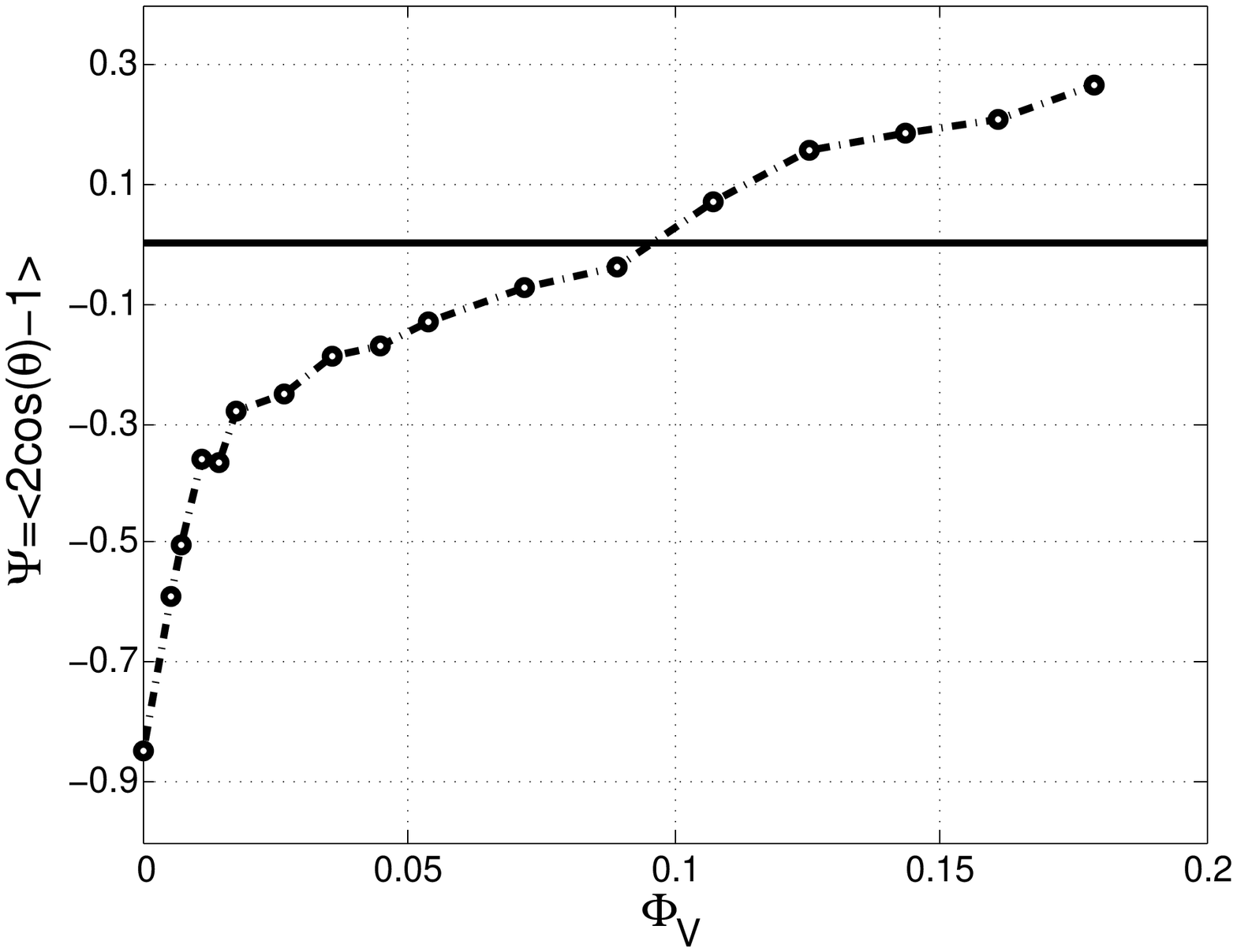,angle=0,width=8cm,height=6cm}
\caption{The top picture shows the distribution function, 
$P(cos(\theta))$ of the mean vertical 
orientation at different volume fraction $\Phi_{V}$. 
The bottom picture shows how the order parameter 
behaves with the volume fraction $\Phi_{V}$. 
The oblate ellipsoid aspect-ratio is $\Delta r=0.4/1.5$, 
the equivalent sphere has $R_{equi}=0.97$ 
and the Reynolds number $Re\approx 7$.}
\label{fig17} 
\end{center}
\end{figure}  

Figure \ref{fig17} (top) presents the vertical 
distribution function, $P(\cos(\theta))$ at 
moderate Reynolds number. For all volume 
fractions, $P(\cos(\theta))$ 
presents a larger distribution around 
$cos(\theta)\approx 0$ ($\theta\approx 90^{o}$), which 
tends to be much flatter ($cos(\theta)\geq 0.15$) 
than in fig. \ref{fig10}.   
For the other angular variables, $\phi$ and $\psi$, the distributions show a peak around 
$\cos(\phi)\approx 0$, $\cos(\psi)\approx 0$, and for larger volume fractions, they follow a
constant behavior. 

The bottom of fig. \ref{fig17} shows the behavior of the 
orientational parameter $\Psi$ at moderate 
Reynolds number. For one oblate ellipsoid 
($\Phi_{V}\rightarrow 0$), the value of $\Psi$ is closer to $-1$  
(vertical alignment), as in the case of 
low Reynolds number fig. \ref{fig12}.
The intermediate maximum is not seen 
in figure \ref{fig15}, and the point at which the 
orientational parameter $\Psi$ vanishes, is shifted 
slightly to the right ($\Phi_{V}\approx 0.1$)  
fig. \ref{fig17} (bottom). This shift in $\Psi$ is 
also seen in the case of fibers, when the Reynolds 
number increases, by a factor of 5, \cite{Esaprl}.  

\section{Outlook and Conclusions}

We have simulated the sedimentation of oblate 
ellipsoids at small volume fraction ($\Phi_{V}\leq 0.2$) 
and small Reynolds number ($Re\approx 10^{-2}$). 
We have found that at intermediate volume fraction 
the settling velocity exhibits a local maximum 
which to our knowledge has never been reported in the 
literature. It would be desirable to experimentally verify this maximum. 

This local maximum in the velocity can be 
related to the non monotonic behavior of the vertical 
orientation of the oblate ellipsoids along gravity, 
which is shown in figures \ref{fig9}, and 
\ref{fig10},  and can be explained by the 
``cluster'' formation shown in fig. \ref{fig1}, which is 
also found in fiber-like suspensions \cite{Herzhaft}. 

At low Reynolds number the orientational order 
parameter $\Psi$ vanishes around $\Phi_{V}\approx0.08$ 
fig. \ref{fig9}. 
As $\Phi_{V}$ decreases the orientational alignment 
with gravity increases as shown in fig. \ref{fig9} and 
\ref{fig17} (bottom), as for low and moderate Reynolds number 
and in the limit $\Phi_{V}\rightarrow 0$ 
a single ellipsoid aligns with gravity, which is a 
distinctive feature of the steady-state 
regime for a single oblate ellipsoid as reported in references
 \cite{FonsH} and \cite{Galdi}.

We also present data at moderate Reynolds number 
($Re\approx 7$) for the sedimentation velocity of 
oblate ellipsoids as the volume fraction $\Phi_{V}$ 
is increased. As in the case of low Reynolds 
number the ellipsoids have a smaller 
sedimentation velocity than the equivalent spheres, fig.
\ref{fig2} and \ref{fig16}. The data for 
ellipsoids and spheres follow the Richardson-Zaki 
law \cite{RiZaki} with exponents 
($n_{Ellip}\approx 3.2$,$Re= 10^{-2}$) and ($n_{Sph}\approx 4.0$,$Re=7$) 
respectively.
The $P(\cos(\theta))$ distribution presents 
a larger alignment of ellipsoids 
with gravity compared to those with small Reynolds number. 
The vanishing of the order parameter is slightly shifted 
($\Phi_{V}\approx 0.1$) to the right as the Reynolds number 
increases (see fig \ref{fig17}, bottom). 
The alignment with gravity is present for small and moderate 
Reynolds number as $\Phi_{V}\rightarrow 0$, as shown in fig. \ref{fig17} 
(top) and fig. \ref{fig10}, which is in agreement with the orientational 
behavior of a single ellipsoid \cite{FonsH}. All the 
simulations in this work are located in the steady-falling regime of our 
previous work \cite{FonsH} for a single oblate ellipsoid.       

One open problem that should be addressed in the future 
is the behaviour of 
velocity fluctuations as the container system size 
and the ellipsoid aspect-ratio 
increases.
This work has been carried out for sedimenting spheres
but not for ellipsoids, \cite{Ramas}.

\section{Acknowledgments}

This research is part of the SFB-404-Project A7. 
F. Fonseca thanks for helpful discussions with S. 
Schwarzer, E. Kuusela, T. Ihle and R. Vasanthi.

\end{document}